\documentclass{article}
\usepackage{graphicx} % Required for inserting images
\usepackage{subfiles}
\usepackage{multicol}
\usepackage{float}
\usepackage{hyperref} 
\usepackage{fontawesome}
\usepackage{enumitem}
\usepackage{caption}
\usepackage{subcaption}

\title{Music-triggered fashion design: from songs to the metaverse}
\author{Martina Delgado, Marta Llopart, Eva Sarabia, \\ Sandra Taboada, Pol Vierge, Fernando Vilariño, Joan Moya Kohler,\\ Julieta Grimberg Golijov, Matías Bilkis \\ \\
 Computer Vision Center, Barcelona, Spain. \\
  Universitat Autònoma de Barcelona, Barcelona, Spain.}

\date{}

\begin{document}

\maketitle
\begin{abstract}
    The advent of increasingly-growing virtual realities poses unprecedented opportunities and challenges to different societies. Artistic collectives are not an exception, and we here aim to put special attention into musicians. Compositions, lyrics and even show-advertisements are constituents of a message that artists transmit about their reality. As such, artistic creations are ultimately linked to feelings and emotions, with aesthetics playing a crucial role when it comes to transmit artist's intentions. In this context, we here analyze how virtual realities can help to broaden the opportunities for musicians to bridge with their audiences, by devising a dynamical fashion-design recommendation system inspired by sound stimulus. We present our first steps towards re-defining musical experiences in the metaverse, opening up alternative opportunities for artists to connect both with real and virtual (\textit{e.g.} machine-learning agents operating in the metaverse) in potentially broader ways. 
\end{abstract}

\section{Introduction}
Music is uniquely experienced by individuals, but it is often unified by the artist's vision. This vision reflects an artistic understanding of reality, which is not only transmitted by the musical piece but also by other means, \textit{e.g.} visual representations such as album covers, scenographies or even dress designs for live-performances.

\vspace{1.5mm}

More generally speaking, aesthetics plays a crucial role in conveying the mood and emotions that artists seek to transmit with a creation, \textit{e.g.} a musical piece. While sound stimulus on its own strongly reverberate with the hearer (\textit{e.g.} think on different moments in song), message intention can clearly be complemented by using available visual resources (\textit{e.g.}  fireworks or smoke in a live performance).

\vspace{1.5mm}

On the other hand, virtual realities are increasingly gaining terrain in our daily lives, and art is not excepted to this fact. In turn, artists are beginning to utilise tools such as the metaverse for concerts. Virtual art expositions and even fashions shows are only more examples of physical events that are shifting towards virtual scenarios. It is then of utmost importance to develop novel tools to get profit of opportunities offered by such venues nature, as well to address different challenges and social implications of such a paradigm shift. 

\vspace{1.5mm}

Here, we provide first steps towards analyzing the role that virtual realities can play when it comes to artist's aesthetics, with the goal of conveying a message towards the audience.

\vspace{1.5mm}

Events happening in the virtual realm clearly lack the particular experience of live performances, \textit{e.g.} the experience of a musical performance is highly richer in real life, rather than broadcasted in the metaverse. shiftThis raises the need of bringing different parties closer, \textit{e.g.} performers and audience, in such virtual experiences. Firstly, exploiting the possibilities offered by virtual realities might make (virtual) artistic performances richer and more enjoyable. Secondly, enhancing the connection between the performer with the audience is crucial to guarantee the transmittance of artist's message. In this respect, the aesthetic scope can be thought to broaden, as novel technologies such as generative models can readily be implemented in order to compensate the lack of physicality. 

\vspace{1.5mm}

An overlooked aesthetic aspect in the virtual reality is that of cloth design, and is strongly linked to live performances. Most approaches rely on personalizing cloth shapes, with the base outfit being static. However, music dynamics can easily be linked to the outfit of either performer, attendant or both in a virtual reality. While music-triggered cloth changes are practically impossible in real life, getting profit of specific climax in a song can serve as a potential new bridge between the musician and her audience. 

\vspace{1.5mm}

\begin{figure}[t!] % Usar H para forzar la posición exacta
    \centering
    \includegraphics[width=1.\textwidth]{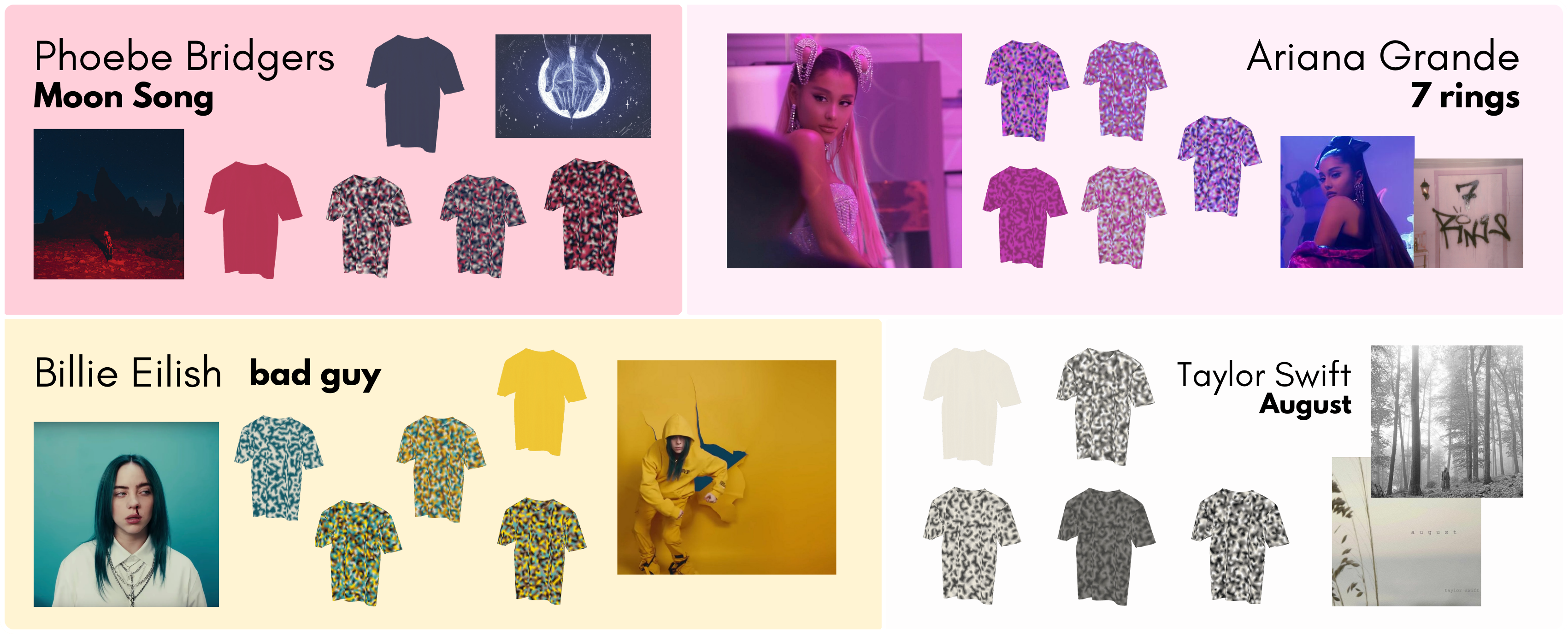}	
    \caption{\textbf{Music-triggered recommendation system}. We show several designs obtained by our recommendation system, each associated to a different songs sample. Here, song features are extracted and further processed by our model, that imprints visual patterns into an avatar's metaverse dressing code. In this example, we show T-Shirt patterns of different songs based on: Moon Song by Phoebe Bridgers, 7 rings by Ariana Grande, Bad Guy by Billie Eilish and August by Taylor Swift.}
     \label{fig:portada}
\end{figure}

Here we introduce first steps towards a music-triggered recommendation system for fashion design, as captured by Fig.~\ref{fig:portada}. Such is intended to be deployed at the metaverse, and to be used in real-time by interpreting current moods in both the audience and the musical piece being played; moreover our method is conceived as a pattern-retrieving device, whose outcome is to be shown in avatar's dressing codes. Such a method is indented to automatize cohesively a temporal aesthetic conception derived from the song being played, and to perfectly encapsulate the vibe of the song as its creator intended. Interestingly enough, whether this or similar mechanisms can succeed in capturing the essence of artist's feelings encoded in the song, or fail into (a potentially higher) ambiguity, is a matter of debate. In turn, we show evidence that even basic approaches linking colours to musical genres fail to encapsulate targeted aesthetic, being ambiguity at the core of the question. Whether state of the art machine learning methods deployerd in virtual realities can serve as a stepping stone towards complementing emotion communication by music, or this is merely an entertainment application inducing artificial and unnecessary ambiguities, is certainly an interesting debate, which we partially address in Sec.~\ref{sec:discu}. 

\section{Bridging music and fashion design}
In the following we describe our music-triggered recommendation system which retrieves cloth design fashion colour patterns. As such, it is indented as a method to capture the essence of artwork's aesthetics, and we here take first steps towards automatizing this task. We remark that while more sophisticated versions of the different components are to be deployed, the value of this contribution resides in both framing and analyzing the role of extended aesthetics in virtual realities to convey artist's messages.

\vspace{1.5mm}

\begin{figure}[t!]
    \centering 
    \includegraphics[width=1.\textwidth]{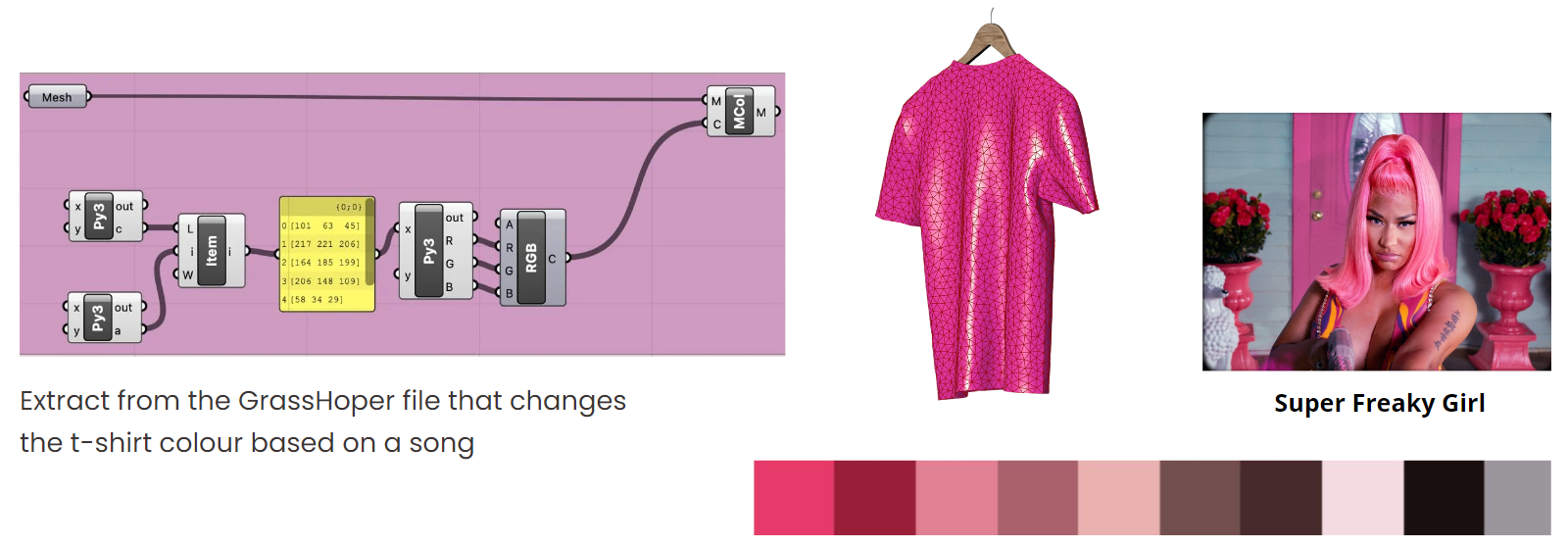}	
    \caption{\textbf{Recommendation system workflow with Rhinoceros generation of Super Freaky Girl}. Grasshoppers interface is shown on the left, linking the colour palette with the song. On the right, the first image appearing on the google search \textit{Super Freaky Girl Nicki Minaj} is shown, along with the generated colour-palette and the T-shirt example.}
    \label{fig:rhino}%
\end{figure}

Our implementation incorporates a palette generation algorithm based on image web scraping with 3D model integration for pattern design, and visualized on a standard T-shirt. A sketch representation of the overall workflow is shown in Fig.~\ref{fig:rhino}, and in the following we explain each of the modules in detail.

\vspace{1.5mm}

\textit{Colour palette and database generation}. A subset of 9K popular songs have been selected from the Spotify API~\cite{spotifySpotifyDevelopers}, in order to create a database with song data and colours sets. The image web scraping algorithm consists of a Google search with a defined \textit{song + artist} structure, and thereby storing the first five appearing images, understood as the most representative ones. The popularity threshold for song selection has been set to 50 (this is an internal score of Spotify API) to ensure that the web scraping yields representative pictures. We remark that this raises up a big concern of our method, which leaves non-mainstream artists aside due to a lack Google images visibility, an issue that we aim to adress in the near future. Following, pixel extraction and $k-$mean clustering have been applied to retrieve a 10 colour palette ($k = 10$) standing for a (simplified representation of) song's aesthetic, based on cluster centres. The generated palettes are thus linked to each song and added to the database.

\vspace{1.5mm}

\textit{Cloth-design pattern generation}. The generative design process for creating patterns is Rhino~\cite{mcneel-rhinoceros}, with strong usage of its 3D modelling-tool plugin, Grasshopper. These parametric design tools are crucial in transforming features that are captured by the recommendation system into tangible patterns. As such, they stand out as powerful platforms for computational modelling, offering designers, architects and artists a versatile toolkit to explore complex geometries and optimise designs. 

\vspace{1.5mm}

The workflow of our design generation begins with a search on the previously-created songs database (web-scrapped database) is conducted based on the title of the song. If found, we extract the colour palette and randomly decide the number of colours appearing in the palette associated to that songs. Hence,the extracted RGB values are then passed as colour parameters to the clothing mesh to visualise the final cloth design (T-shirt in the examples shown here).

\vspace{1.5mm}

Having detailed our initial implementation of a generative colour-based pattern recommendation out of musical stimulus, we now turn to discuss potential implications in the virtual societies inhabiting the metaverse ecosystem.

\section{Aesthetics, common sense and ambiguity}\label{sec:discu}
Let us now discuss social implications of a potentially artist-to-audience communication enhancer device, such as the one introduced above.

\vspace{1.5mm}

While we have restricted to a colour-palette being decided by pre-defined song features, we readily note that alternative song characteristics are also to be considered in the future, such as tempo, musical instruments timbre, lyrics among many others. With this, we aim to highlight that the constituents of aesthetics beneath a musical piece are perhaps not even a countable set, since the relationship between emotions, reflections and artworks are often highly complex and difficult to capture in a conceptual way.

\vspace{1.5mm}

While we have restricted to a colour-palette being decided by pre-defined song features, we readily note that alternative song characteristics are also to be considered in the future, such as tempo, musical instruments timbre, lyrics among many others. With this, we aim to highlight that the constituents of aesthetics beneath a musical piece are perhaps not even a countable set, since the relationship between emotions, reflections and artworks are often highly complex and difficult to capture in a conceptual way.

\begin{figure}[t!]
    \centering 
    \includegraphics[width=1.\textwidth]{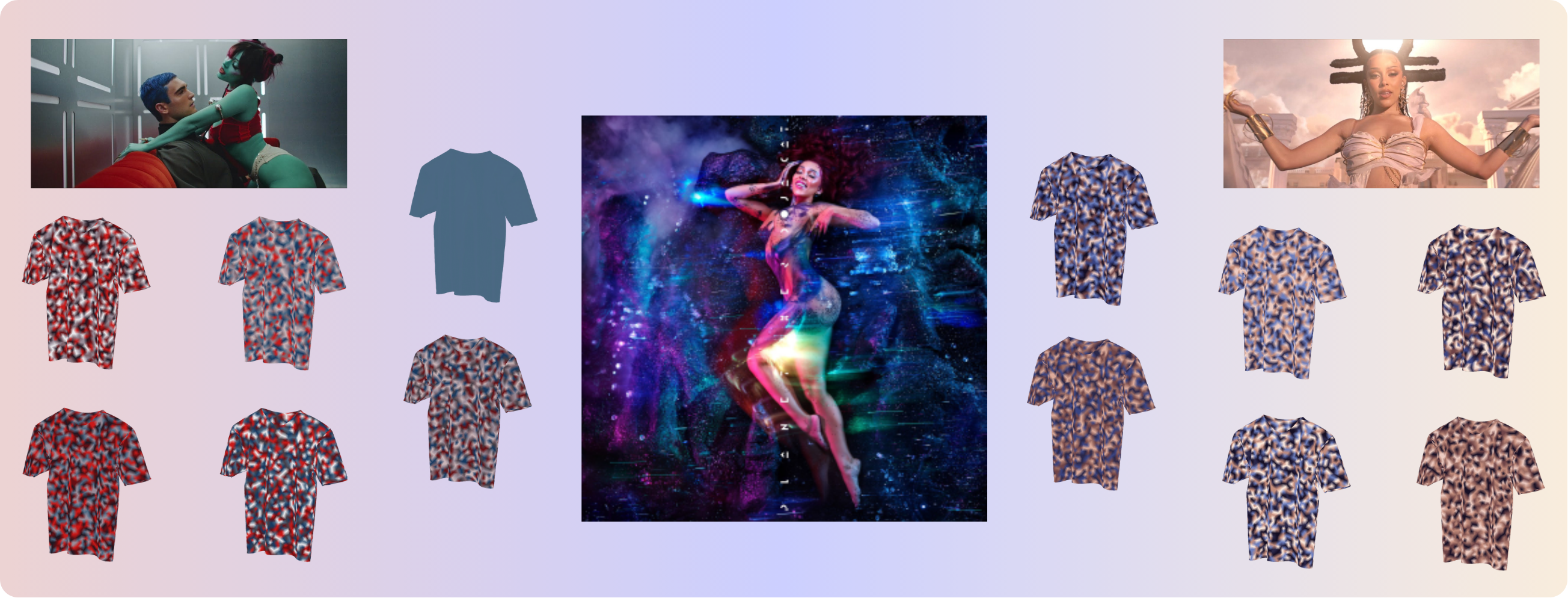}	
    \caption{\fontsize{9}{9}\selectfont\textbf{Generated T-shirts from songs from the same album. }{Comparison between Need to Know by Doja Cat (left) and You Right by Doja Cat featuring The Weeknd (right), both from Doja Cat's third studio album Planet Her (centre).}}
    \label{fig:nuances}
\end{figure}

In this sense, capturing artwork aesthetics with a colour palette is clearly not a conclusive task. For example, while songs from the same album tend to share similar aesthetics, they do so with different feature weights. This induces a notion of ambiguity in how our model captures the aesthetics of an artwork. For instance, our method seems to handle nuances between songs coming from the same album, as shown in Fig.~\ref{fig:nuances}. Here, we show a comparison of patterns obtained by different songs of Planet Her (an album from Doja Cat's), shedding light into the fact that capturing artwork aesthetics is a highly subjective task.  
 
\setlength{\parskip}{0mm}

\vspace{1.5mm}

Such an ambiguous notion on aesthetic quantification has a counter-side on how biases are constructed, both by virtual and real agents. In turn, a key issue that we encountered is the inability of our model to capture the underlying patterns between song features and aesthetic colour palettes. This failure highlights the strong influence of the artist's vision on its music and suggests that preconceived ideas about how a song or genre should look might not be as reality-based as commonly believed.
Moreover we remark that our recommendation system is trained at the level of songs, and not albums. While our implementation is arguably simple, it is not obvious that large-scale learning models trained on massive web-scrapped artworks data would succeed in this tasks; we note that we failed in training deep generative architectures under this premise, precisely because the webscrapped data was both ambiguous and scarse.
\vspace{1.5mm}

\begin{figure}[t!]
    \centering
    \begin{subfigure}[b]{0.49\textwidth} % Adjusted width
        \centering
        \includegraphics[width=\textwidth]{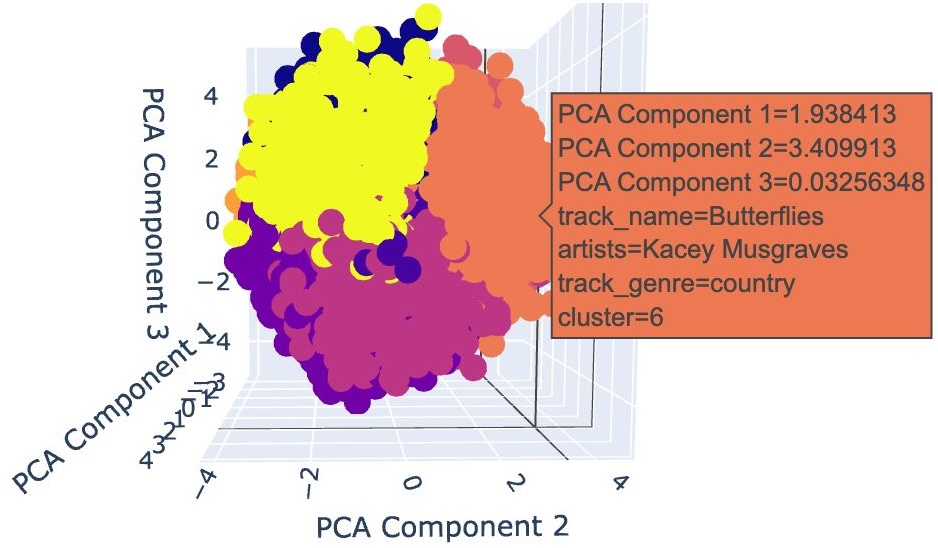}
        \caption{}
        \label{fig_sub1}
    \end{subfigure}
    \hfill % This command ensures subfigures are placed next to each other
    \begin{subfigure}[b]{0.49\textwidth} % Adjusted width
        \centering
        \includegraphics[width=\textwidth]{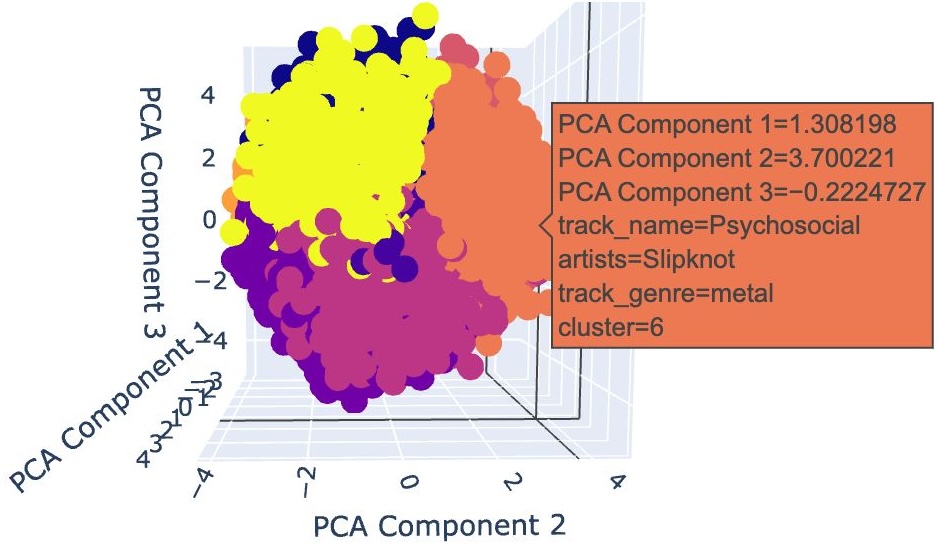}
        \caption{}
        \label{fig_sub2}
    \end{subfigure}
    \caption{\fontsize{9}{10}\selectfont\textbf{Clusters of songs based on colour palettes.} {Comparison between Butterflies by Kacey Musgraves (a) and Psychosocial by Slipknot (b).}} % Optional overall caption
    \label{fig:clustering}
\end{figure}

As a matter of fact, genres are often associated with certain colours, such as pop music being linked with bright colours like pink. However, these preconceptions can easily be challenged. The model's struggle to consistently link song features with specific colours indicates that genre-colour associations might not be as universally valid. This can be evidened, for instance, by performing cluster analysis on the data (see Fig.~\ref{fig:clustering}), which shows that colour associations varied significantly even within the same genre. This observation indicates that the unique synesthesia each artist experiences with music makes it difficult, if not impossible, for a model to extrapolate consistent patterns. This fact points to a bias in the collective perception of colour association in music, which is ultimately linked to cultural and collective constructions. Whether virtual societies can build different pre-conceptions of such associations, and via which new tools or technologies, is an intriguing line of research.%$$ discovery%While this phenomena is a matter of not deploying a sufficiently complex technology, or is an intrinsic ambiguity of artworks and emotion expression, is a matter of discussion.

\vspace{1.5mm}

On the other hand, while we have presented the recommendation device as an artist-to-audience communication channel, we might easily find different use cases. In turn, an interesting research direction is that of building stronger bonds beneath virtual communities, and whether a collective generation of fashion designs might be feasible by deploying future versions of the here-presented recommendation system. This is particularly important in the context of live performance and real-time audience-to-artist feedback, where the artist is influenced by its audience and vice versa; while this phenomena often happens in the physical world, a plethora of unexplored opportunities does open in virtual realities. In this line, it is necessary to understand whether a common sense can be constructed in a virtual society, and whether machine-learning technologies can help to elucidate it (for instance, by extracting relevant common features about different individuals). Finally, the question of whether a common sense about understanding and feeling an artwork beneath an hybrid human-machine society inhabiting the metaverse can be framed or not does also raise up.

\vspace{1.5mm}

\vspace{1.5mm}

We aim to finish this contribution with some open questions. By introducing this music-triggered fashion recommendation system, some aspects regarding human-machine creativity are ought. Among them, questioning the nature of live music and the authenticity of concerts is almost unavoidable. Key questions include:

\vspace{1.5mm}

\textit{Redefining Live Music}. What constitutes live music in the context of increasing technological integration? Our device challenges traditional notions of live performances by potentially blending virtual elements with physical presence, prompting a reevaluation of what do audiences value in live-music performances.

\vspace{1.5mm}

\textit{Authenticity of Virtual Concerts}. Can virtual concerts be considered less authentic when they lack the proximity and physical presence of the artist? This question delves into the essence of concert experiences, questioning whether virtual interactions can replicate the emotional and sensory engagement of being physically present at a live performance.

\vspace{1.5mm}

\section{Related Work}
State of the art regarding music and fashion AI-powered devices is scarce, we refer nevertheless to a recent prospective review about Fashion intelligence in the Metaverse in ~\cite{Mu2024}. Moreover, Ref.~\cite{breese2020} provides a comprehensive approach to the live music performance from IoT. Notwithstanding, a project sharing motivation with ours is Emotional Clothing, a clothing collection by Polish designer Weglinska composed of colour-changing clothing based on body temperature, stress levels, movements, and even motions. The work shows a garment that changed colour with every laugh, conversation, scream or whisper, or each time they made contact with skin. Departing from this, our approach focuses on colour changes on clothing depending on music, with future perspective on a full outfit generation device.

\vspace{1.5mm}

\section{How this work was conceived?}
This project is a continuation of a final project in the Social Innovation course of the AI degree at Autonomous University of Barcelona, held form January-July 2024. Such a course, whose modality is innovative on its own, consists on splitting the students into small working groups, each developing a project. In this case, the project was related to meta-verse technologies, and the goal of the course is to evaluate the social impact of the prototypes developed during it. For more information about this education modality, we refer to Ref.~\cite{BILKIS2024CHA}.   

\section{Code availability}
Our implementation of the  our music-triggered recommendation system for cloth patterns is made publicly available in the GitHub repository
\href{github.com/martiinsssssss/metaverse-clothing}{\faGithub\ martiinsssssss/metaverse-clothing}. Note that a demo video can be found there as well.

\section{Acknoledgments}
This work was done partially under the support of the project Cátedra UAB-Cruilla, funded by the Spanish Governement (TSI-100929-2023-2 (Prov)). All our acknowledgement to the experts participating in the interviews and mentoring during the SI course at UAB, among them Carla Miernau,  Andrea Ross, Abril Riera, Aloma Martí, Laia Torras.

\small
%\begin{comment}
%\bibliographystyle{ieeetr}

%\bibliography{library}
%\end{comment}

\end{document}